\documentstyle[11pt,AATS,epsf]{article}	

\begin{document}	


\title{The Age of the Galactic Bulge $^{1,2}$} 


\author{R. Michael Rich}
\affil{Department of Physics and Astronomy, University of
California at Los Angeles, Los Angeles, CA, USA}


\begin{abstract}
The dominant stellar population of the central bulge of the Milky Way is
old, with roughly solar metallicity.  The age is very similar to that of
the old metal rich bulge globular clusters and to 47 Tucanae, which has an
age of 13$\;$Gyr.  Stellar composition measurements from Keck/HIRES
confirm that bulge stars are enhanced in Mg and Ti.  New HST/NICMOS data
are consistent with an old stellar population dominating the central
100$\;$pc of the Milky Way.

New infrared photometry has been reported for the bulge of M 31.  Although
bright asymptotic giant branch stars are observed in the infrared, the
data are most consistent with an old stellar population.  
\end{abstract}

\footnotetext[1]{Based on observations obtained at the W. M. Keck
Observatory, which is operated jointly by the California Institute of
Technology and the University of California.}
\footnotetext[2]{Based on observations with the NASA/ESA Hubble Space
Telescope, obtained at the Space Telescope Science Institute, which is
operated by the Association of Universities for Research in Astronomy,
Inc, under NASA contract NAS 5-26555.}


\section{Introduction}

Our description of galaxy evolution relies on a balance of data from high
redshift, and detailed population studies of the fossil record.  Ages,
composition, kinematics, and structure of stars preserve evolution in the
greatest detail, but are available only for the nearest galaxies.  The
advent of 8--10m class telescopes and HST marks one of those rare periods
where technology conspires to deliver dramatic advances at high redshift.
At such times, attention naturally turns away from study of the fossil
record, despite the dramatically tighter constraints such work provides.
When uncertainties mount in the high-redshift work (as they often do)
observers return to the fossil record to resolve the ambiguity.  Detailed
population studies can give relative ages to accuracies of 1--2$\;$Gyr,
and detailed composition measurements that are impossible using integrated
light.

\subsection{What is the Bulge?}

The central bulge of the Milky Way has not easily won recognition as a
distinct stellar population.  Until the dramatic images obtained by the
{\sl COBE} satellite, showing a clear bulge in the infrared, the
possibility could not be ruled out that our central population is instead
an extension of the thick disk, inner halo, or perhaps even an
intermediate age bar.   From the infrared imagery, microlensing studies,
and stellar dynamics, we now have good constraints on the total mass of
the bulge, of order $2 \times 10^{10}\;M_\odot$ and a self-consistent
dynamical model (Zhao, Rich, \& Spergel 1996).  The typical velocity
dispersion of the stars is $\sim110\; \rm km\;sec^{-1}$; the mean abundance of the
bulge is $\sim-0.3\;$dex (McWilliam \& Rich 1994) and the abundance
distribution function is consistent with the closed box chemical evolution
model (Rich 1990).

The evolved stars in the bulge are quite different from those found in the
typical globular cluster.  The well known RR Lyrae stars first identified
by Baade are present and represent the metal-poor population.  However,
the bulge has a range in abundance and the most metal rich giants in the
bulge exceed the solar abundance.  The horizontal branch in the field
color--magnitude diagram (CMD) is dominated by red horizontal branch (HB)
(clump) stars, not by the bluer RR Lyrae stars, and in optical colors, the
tip of the red giant branch is observed to descend fainter than the red
clump luminosity due to blanketing in the M giants (Rich et al.\ 1998).
The asymptotic giant branch (second ascent) stars are late M giants and
can reach $M_{\rm bol} = -5$.  The presence of such luminous stars in the
bulge fueled well-justified speculation that a widespread intermediate age
population must be present there.  We now know that stars in the 13$\;$Gyr
globular cluster age range can reach that luminosity, if they are metal
rich (Guarnieri, Renzini, \& Ortolani 1997).

Challenged by heavy and variable reddening, extreme image crowding, the
spatial depth of the population, and the broad range of metallicity,
observers have found it challenging to measure the age.  Arp (1965)
produced a color--magnitude diagram for the giants in Baade's Window from
photographic $BV$ photometry.  Van den Bergh (1972) refined further
reddening and distance estimates, and argued for a metal-rich giant branch
in his photometry of Baade's Window.  Blanco's $RI$ photographic plates
on the newly commissioned 4m telescope at CTIO led to a breakthrough: for
the first time the giant branch was clearly defined, permitting a sample
of K giants to be selected for spectroscopy (Whitford \& Rich 1983).  Van
den Bergh \& Herbst (1974) developed some age constraints for the Plaut
field ($8\deg S$) of the nucleus.  Terndrup (1988) attained an old main
sequence turnoff in this field, but could not resolve the metal rich bulge
population deep in Baade's Window.  The debate concerning the age range in
the bulge remains active to this day.

Related to the age is the formation time scale (review in Rich 1999).
The time scale for the bulk of star formation can be constrained by
modeling the turnoff.  One would like to use the asymptotic giant branch
(AGB) to constrain residual star formation on Gyr time scales, but the
lack of useful correlation between age and tip luminosity for metal rich
stars poses a problem.  Composition (see Section 3, below) offers some
interesting constraint on the formation time scale.  To date, the best
constraints argue that the bulge is globular cluster age, and formed in
$<1\;$Gyr.

\subsection{Historical Perspective}
  
The importance of the bulge, and the description of bulge populations, was
well appreciated by the pioneers in the field, especially Walter Baade.
Although the actual Galactic center is obscured by tens of magnitudes of
extinction, the circumstantial evidence for the nucleus lying toward
Sagittarius was strong enough that Baade searched for RR Lyrae stars in
that direction.  His discovery of a sharply peaked magnitude histogram for
the RR Lyrae stars marked the discovery of the bulge as a stellar
population (Baade 1951).

By the time of the pivotal 1958 Vatican meeting on stellar populations,
the present-day view concerning the age and metallicity of the bulge had
been developed and was correct.  The presence of RR Lyrae stars argued for
an old population, but the presence of late-type M giants (discovered 
in grism surveys by Victor Blanco) argued for high metallicity.  In a
final table summarizing the characteristics of stellar populations, the
Galactic nucleus was classified as old and metal rich, but some lingering
doubt was expressed as to whether the nucleus (bulge) population was as
old as the globular clusters.  This confluence of thinking was summarized
in a prescient and sophisticated analysis of the formation history of the
M 31 bulge, the template stellar population for Population II (Baade
1963):  

``We must conclude, then, in the central region of the
Andromeda Nebula we have a metal-poor Population II, which reaches $-3^m$
for the brightest stars, and that underlying it there is a very much
denser sheet of old stars, probably something like those in M 67 or NGC
6752.  We can be certain that these are enriched stars, because the
cyanogen bands are strong, and so the metal/hydrogen ratio is very much
closer to what we observe in the Sun and in the present interstellar
medium than to what is observed for Population II.  And the process of
enrichment probably has taken very little time.  After the first
generation of stars has formed, we can hardly speak of a `generation',
because the enrichment takes place so soon, and there is probably very
little time difference.  So the CN giants that contribute most of the
light in the nuclear regions of the Nebula must also be called old stars;
they are not young.''

High resolution spectroscopy was, of course, lacking, and the realization
that some globular clusters are very metal poor, and 1/100 solar
abundance, also was absent (witness the mention of M 67 and NGC 6752 in the
same breath).  But the Vatican conference mentions both large age and high
metal abundance for the bulge population.  For approximately the next 20
years, the bulge would often be described as metal poor, despite the very
clear evidence of a wider abundance range, accumulated prior to that
time.  Ultimately, the observational definition of the globular cluster
abundance scale, combined with new work on the bulge, would modify this
erroneous picture.

\subsection{The Relationship to High Redshift Studies}

Although much can be inferred from the evolution of spiral galaxies as a
function of redshift, the well-determined age and age range of one single
bulge population (that of the Milky Way) is a powerful constraint.

The classic bulge formation model is that of Eggen, Lynden-Bell, \&
Sandage (1962) in which bulges form from dissipationless collapse very
early on.  The widely accepted cold dark matter models have some
variations, but basically all concur that elliptical galaxies form from
merged disk galaxies (Kauffmann, White, \& Guiderdoni 1993; Baugh et
al.\ 1998).  At any given redshift, bulges should then be older than
ellipticals.   Some bulges may be related to bars, which could vertically
thicken due to scattering of resonant orbits off of the bar, or due to the
dissolution of bars (Combes 2000, and references therein).  Bulges with
exponential profiles may be more closely related to disks (Carollo et
al.\ 2001).

The most distant normal star forming galaxies are the Lyman-break galaxies
(Steidel et al.\ 1996) which are strongly clustered (Adelberger et al.\
1998; Giavalisco \& Dickinson 2001).  These galaxies have strong outflows
and are evidently metal rich; it is therefore reasonable to suppose that
they will evolve into spheroids.  The connection between the galaxies at
$z\sim 3$ and the population of galaxies that begin to fall in well
defined Hubble types by $z\sim 1$ requires much effort and lies ahead of
us.   By $z\sim 1$ we can begin to classify bulges of spirals, and
ellipticals and galaxies are well enough resolved to even investigate
their star formation histories pixel by pixel.  Ellis, Abraham, \&
Dickinson (2001) studied the evolution of bulges and ellipticals in the
Hubble Deep Fields and found that at any given redshift, ellipticals are
redder than bulges.  Issues such as disk contamination and reddening have
rigorously been accounted for.  Numerous studies have found evidence for
very red ellipticals, and clustering of ellipticals, in the $z\sim 1-1.5$
range (see Stockton's review, these proceedings).  Locally, Peletier et
al.\ (1999) and numerous other studies argue that spiral bulges are as old
as present-day Coma ellipticals.  In the section that follows, I argue
that the stellar populations of the Galactic bulge are consistent with a
large age.  How do we reconcile these findings with the high redshift
studies?  Are most bulges subjected to brief starburst events, due to the
availability of gas from the disk or environment, or are elliptical
galaxies systematically more metal rich?  We have to resolve this tension
between the high redshift data and the fossil record.

\section{Age Constraints From the Turnoff}

Because of the high reddening, uncertainty in the distance modulus, and
presence of foreground stars, one cannot constrain the age of the bulge
with confidence by placing isochrones on the CMD.  The foreground main
sequence disk stars have proven a vexing population, as they overlay the
old main sequence turnoff point precisely and appear as an intermediate
age population.  To increase the contrast between the bulge population and
the foreground, one needs to study fields close to the Galactic center,
where reddening and crowding become serious.  Estimates from modeling and
direct starcounts place the foreground contamination at 10--15\% (see the
CMDs by Feltzing \& Gilmore 2000).

The strongest arguments in favor of an intermediate age population in the
bulge were based on the presence of luminous AGB stars.  An empirical
correlation between AGB star luminosity and intermediate populations was
established using Magellanic Clouds clusters (Aaronson \& Mould 1985).
The $\sim1\;$mag extension at the turnoff point prevented a convincing
resolution of the problem.

Ortolani et al.\ (1995) found a solution to the problem.  Both the
reddening and distance modulus uncertainties could be eliminated if a
differential comparison is made between the bulge luminosity function and
that of an old globular cluster of comparable metallicity.  The crucial
technique is to force fit the bulge field luminosity to the globular
cluster luminosity function at the point of the red clump, and to examine
the agreement at the turnoff point in detail.  Red clump stars are fueled
by helium core burning, and the core mass (hence luminosity) has almost no
dependence on age and little dependence on metallicity.  The bulge is
concentrated spatially and consequently there is little distance
dispersion.  Therefore, when the red clumps are force fit, it is easily
seen that the turnoff rise of bulge and NGC 6553 are identical.
Therefore, using the well established $\Delta V^{HB}_{TO}$ method of age
determination, we are able to show that the bulge and NGC 6553 have
identical ages.

The next question is, how old is NGC 6553?  The metallicity of the cluster
is in some debate right now.  Barbuy et al.\ (1999) find [Fe/H] = $-0.5$ and
alpha elements up by $+0.5\;$dex relative to scaled solar.  Cohen et al.\
(1999) use Keck/HIRES spectroscopy to find [Fe/H] = $-0.16$ and alpha
elements up by $+0.3\;$dex.  R. M. Rich, L. Origlia, \& S. M. Castro (in preparation) use
infrared spectroscopy to find [Fe/H] = $-0.3$ and alpha elements up by
$+0.3\;$dex.  Regardless of which value one adopts, NGC 6553 is a good
match for the field population of the bulge, and only 0.4--0.5$\;$dex more
metal rich than 47 Tuc.

We have determined a white dwarf distance for 47 Tuc (Zoccali et al.\
2001) one of the nearest of the metal rich disk globular clusters.
The distance modulus translates into a turnoff age of $13 \pm 2.5\;$Gyr.
Given the close correspondence between NGC 6553 and 47 Tuc, I adopt an
identical age for the Galactic bulge.  This makes the bulge of order the
age of the halo.
There has been recent work that solidifies this large age, and
additionally places limits on the time scale of bulge formation.  Feltzing
\& Gilmore (2000) show quantitatively that {\it all\/} of the stars lying
slightly brighter than the old main sequence turnoff in the bulge fields
belong to the foreground.  This is also being confirmed in proper motion
surveys based on HST data (K. Kuijken \& R. M. Rich, in preparation).

\subsection{Implications for Red Globular Cluster Systems}

As a final side note, in many elliptical galaxies with globular cluster
systems, the globular clusters are bimodal in their color distribution.
The bluer clusters tend to be spatially extended, while the red clusters
follow the spheroid, just as is the case in the Milky Way (the bulge
clusters follow the bulge).  It is widely accepted (Ashman \& Zepf 1992)
that the red cluster systems are connected with late mergers.  In the
Milky Way bulge, we have shown (one example) that the red clusters are
old, like those in the halo.  However, I would propose that the red
clusters in ellipticals should be considered to have the age of the old
spheroidal stars, unless proof to the contrary can be developed.

\section{Time Scale Constraints From Abundances}

Considering that this is a meeting that discusses both ages and time
scales, we should address the time scale of formation of the bulge.
Strong constraints have been attained from the photometry discussed in the
last section.  The compositions of stars also have the potential to
constrain the time scale for chemical enrichment.  Massive star supernovae
which explode on $10^6\;$yr time scales have ejecta rich in alpha-capture
elements (O, Mg, Si, etc.), the products of nucleosynthesis in their
hydrostatic burning shells.   Type Ia SNe experience a deflagration in
their explosions, and contribute more iron (but on time scales of
$\sim1\;$Gyr) than the massive star (core detonation) supernovae.   The
general trend is that the alpha/iron ratio declines with increasing [Fe/H]
with the more recently formed stars approaching solar composition.  The
more rapid the enrichment, the higher the iron abundance of stars that
maintain an alpha-enhanced composition.  This paradigm, fully described in
(cf. McWilliam 1997) permits us to make relative comparisons of enrichment
time scales in stellar populations.  Figure 1 shows the abundance trends
for bulge giants measured from spectroscopy on Keck, using the HIRES
echelle (Vogt et al.\ 1994).  A more detailed discussion of recent Keck
results on bulge giants is given in Rich \& McWilliam (2000).

\begin{figure}
\plottwo{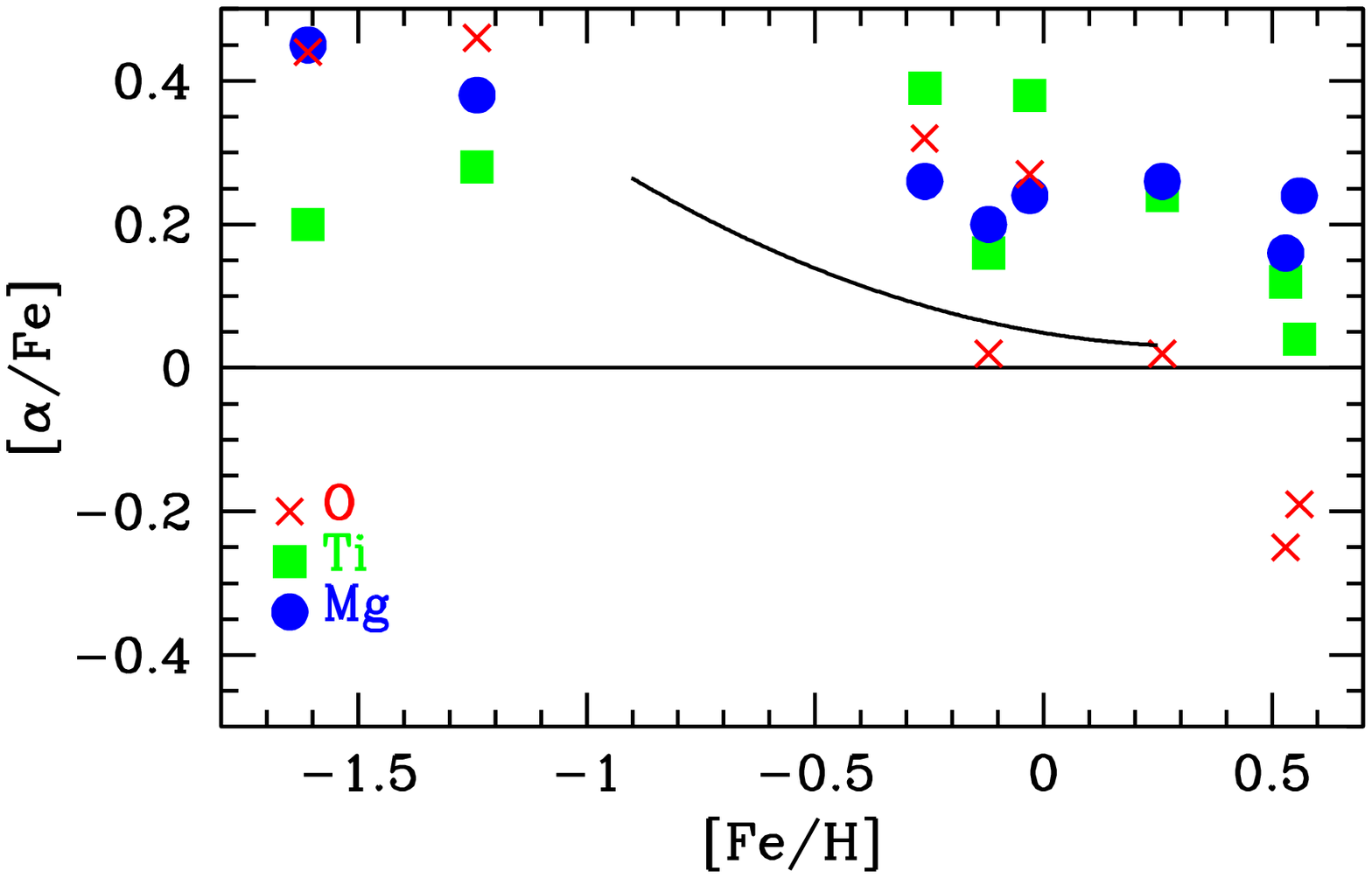}{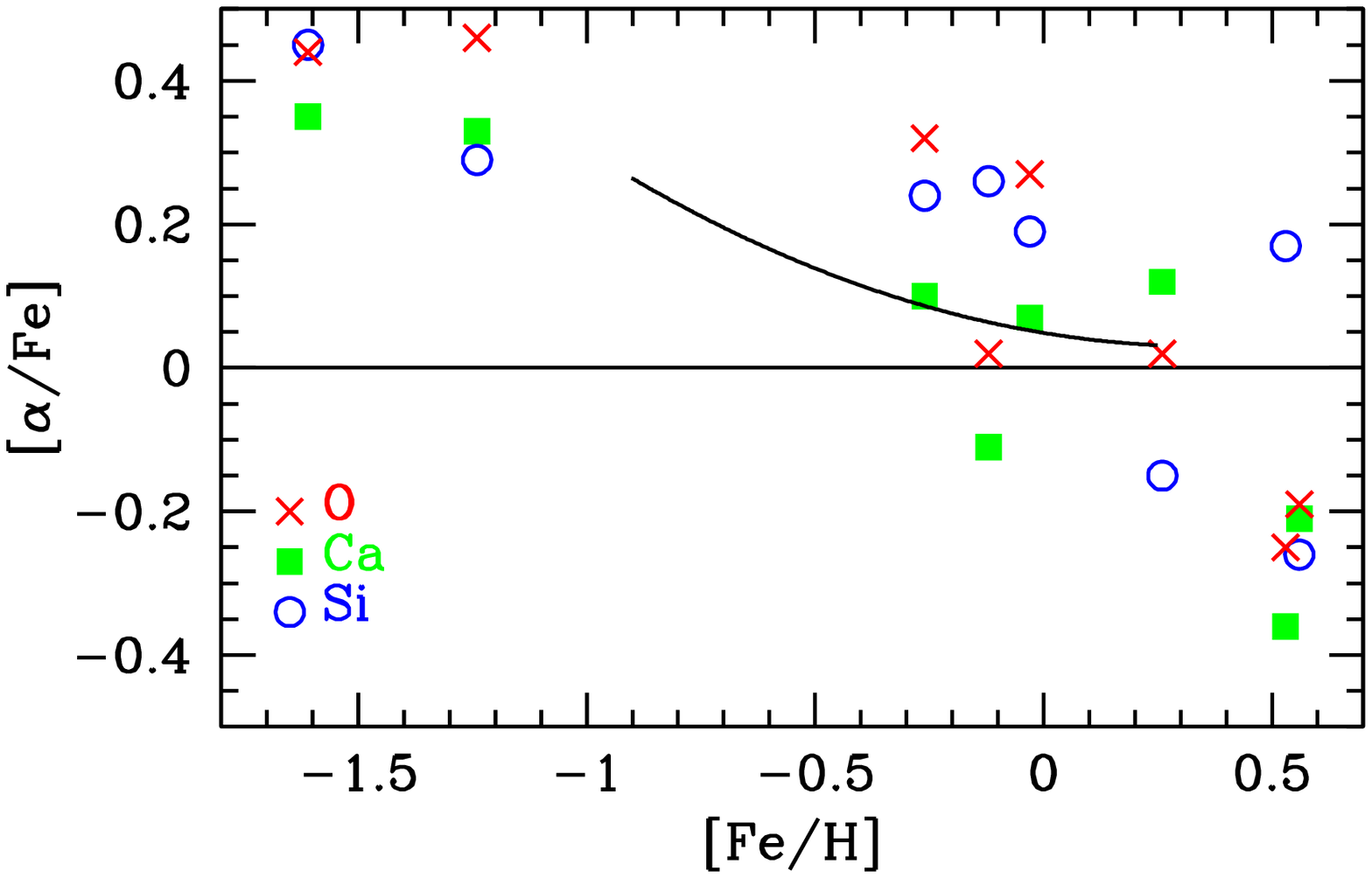}
\caption{Abundance trends for Galactic bulge giants, observed with
Keck/HIRES (Rich \& McWilliam 2000).  The solid line shows the mean
abundance trend for alpha elements in local disk stars (Edvardsson et
al.\ 1993).  Although Mg enhancement is consistent with element production
in Type II SNe, SNe models predict lower Ti production.  The general
enhancement in alpha elements strongly points toward a rapid time scale
for bulge formation.  The enhancements of Mg and Ti relative to Ca and Si
confirm the work of McWilliam \& Rich (1994).}
\end{figure}

The bulge shows a unique pattern of enrichment not shared by any other
Galactic stellar population.  The alpha element Mg remains enhanced even
above the solar iron abundance, as does Ti.  Oxygen follows a less
striking trend, while exceeding the solar abundance at $-0.3\;$dex.  Ca and
Si are modestly enhanced.  This complicated pattern of enhancement is not
predicted by supernova models, but is likely consistent with the bulge
being rapidly enriched by massive stars.   The rapid formation time scale
is consistent with the age constraints from the photometry, and with
observed large star formation rates at high redshift.

\section{The Age of the Nuclear Population}

The stellar population of the Galactic nucleus is one of the most active
and dramatic star forming regions in the Galaxy (Morris \& Serabyn 1996).
Massive star formation dominates the energy input, and includes such
dramatic examples as the $10^6\;L_\odot$ Pistol Star (Figer et al.\ 1998)
and the Arches cluster, with $10^3$ massive O stars and an age of 2$\;$Myr
(Figer et al.\ 1999).  Considering the intensity of star formation near
the nucleus, one expects to find a stellar population with a wide
range of ages, and a present day mass function consistent with a
continuous history of star formation.  Ground-based imaging of the
Galactic Center (Catchpole, Whitelock, \& Glass 1990) found evidence for a
concentration of the most luminous AGB stars toward the Galactic
Center, further strengthening the case that the inner 100$\;$pc of the
Galaxy must have a wide age range.

Using the NICMOS infrared imager on board HST, we obtained deep infrared
photometry of optically obscured fields near the Galactic center (R. M. Rich et
al., in preparation; Figure 2).  Much to our surprise, we found a
clearly-defined red clump and red giant branch, consistent not with an
intermediate age population of ongoing star formation, but rather with an
old stellar population.  We again use the vertical method of age
determination, $\Delta$(mag)$^{\rm HB}_{\rm TO}$, this time comparing our
Galactic Center population to NGC 6553 (Figure 2).  We conclude that the
bulk of the stellar population in these regions is as old as the globular
clusters, just as is the case in the outer bulge.  In light of the star
formation activity in the nuclear region, this is a surprising finding.
The luminosity function and color--magnitude diagram of these fields a
mere 40$\;$pc from the Galactic Nucleus appears identical to that of bulge
globular clusters observed in the same NICMOS bands.  The lack of an
intermediate age population is surprising; perhaps conditions in the
Galactic Center permit only the formation of massive stars, so that
longer-lived low mass stars would not survive.

\begin{figure}
\plotfiddle{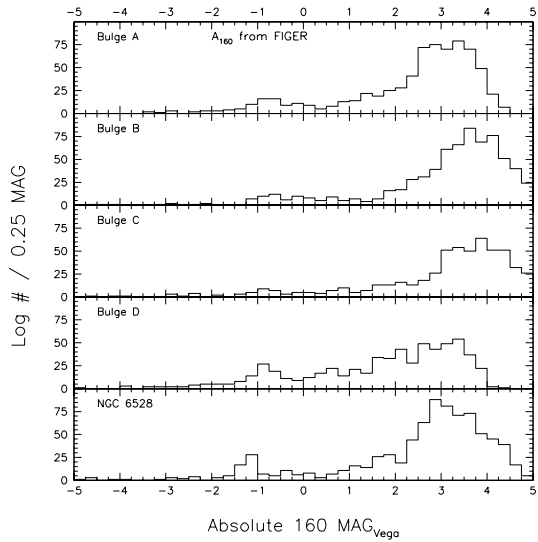}{2.5in}{0}{38}{38}{-130pt}{-70pt}
\caption{$H$-band luminosity functions of the bulge using HST/NICMOS 
(R. M. Rich et al., in preparation).  The age-sensitive gap
between the red clump $(H \sim -1.25)$ and the main sequence rise at $H
\sim 3$ is very similar for the Galactic Center fields (10--40$\;$pc from
the nucleus) and the old metal rich globular cluster NGC 6528 ({\it lower
panel\/}).  The nucleus is dominated by an old population.}

\end{figure}

\section{Conclusion}

Using the widely accepted vertical method of age determination, the
Galactic bulge has approximately the age of the Galactic globular cluster
47 Tuc, $13 \pm 2.5\;$Gyr.  The vertical method (using infrared luminosity
functions) also shows that the population within 10--40$\;$pc of the nucleus is
old.  The composition of the bulge K giants is consistent with early,
rapid formation of the bulge.  These age measurements indicate that the
Galactic bulge/bar is among the oldest Galactic stellar populations.  If
the population formed by the vertical thickening of a massive disk, it
must have done so very early in the Galaxy's formation history.

Local group bulges are also consistent with this picture.  The infrared
luminosity functions of M 32 (Davidge et al.\ 2000) and the bulge of M 31
(Stephens et al.\ 2001) terminate at $M_{\rm bol} = -5$---confirming early
ground-based IR imaging by Rich \& Mould 1991.  While the old globular
cluster NGC 6553 has bright AGB stars, the complicated behavior of metal
rich stars on the asymptotic giant branch probably makes it difficult to
rule out an intermediate age component in these local group bulges.
Better age constraints must come from modeling the color and integrated
spectral energy distribution, to constrain trace intermediate-age main
sequence stars.  Our capability to use the old main sequence turnoff point
to constrain the age of the bulge is a golden opportunity not available
for even the nearest Local Group bulges.  In terms of age, the Galactic
bulge much more resembles the halo than it does the disk, and all the data
point to it being completely in place very early on in the Galaxy's
history.

\acknowledgments
Support for this work was provided by NASA through grants GO-7832,
GO-7465, and GO-7826 from the Space Telescope Science Institute, which is
operated by AURA, Inc., under NASA contract NAS 5-2655.

\begin{references}

\reference Aaronson, M., \& Mould, J. R. 1985, \apj, 288, 551

\reference Adelberger, K. L., Steidel, C. C., Giavalisco, M., Dickinson, M.,
Pettini, M., \& Kellogg, M.  1998, \apj, 505, 18.

\reference Arp, H. C.  1965, \apj, 142, 402

\reference Ashman, K. M., \& Zepf, S. E.  1992, \apj, 384, 50

\reference Baade, W.  1951, Pub.\ Obs.\ Univ.\ Michigan, 10, 7

\reference Baade, W. 1963 in The Evolution of Galaxies and Stellar
Populations (Cambridge: Harvard University Press), 256

\reference Barbuy, B., Renzini, A., Ortolani, S., Bica, E., \& Guarnieri, M. D. 1999,
\aap, 341, 539

\reference Baugh, C. M., Cole, S., Frenk, C. S., \& Lacey, C. G.  1998, \apj, 498, 504

\reference Carollo, C. M., Stiavelli, M., de Zeeuw, P. T., Seigar, M.,
\& Dejhonge, H.  2001, \apj, 546, 216

\reference Catchpole, R. M., Whitelock, P. A., \& Glass, I. S. 1990, \mnras,
247, 479

\reference Cohen, J. G., Gratton, R. G., Behr, B. B., \& Carretta, E.
1999, \apj, 523, 739

\reference Combes, F. 2000, in ASP Conf.\ Ser.\ 197, Dynamics of Galaxies:
From the Early Universe to the Present, ed.\ F. Combes, G. A. Mamon, \&
V. Charmandaris (San Francisco: ASP), 15


\reference Davidge, T. J., Rigaut, F., Chun, M., Brandner, W., Potter, D.,
Northcott, M., \& Graves, J. E.  2000,  \apj, 545, L89

\reference Edvardsson, B., Andersen, J., Gustafsson, B., Lambert, D.,
Nissen, P., \& Tomkin, J.  1993, \aap, 275, 101

\reference Eggen, O. J., Lynden-Bell, D., \& Sandage, A.  1962, \apj, 136, 748

\reference Ellis, R., Abraham, R. G., \& Dickinson, M.  2001, \apj, 551, 111

\reference Feltzing, S., \& Gilmore, R. 2000, \aap, 355, 949

\reference Figer, D. F., Kim, S. S., Morris, M., Serabyn, E., Rich, R. M.,
\& McLean, I. S.  1999, \apj, 525, 750

\reference Figer, D. F., Najarro, F., Morris, M., McLean, I. S., Geballe, T. R., 
Ghez, A. M., \& Langer, N.  1998, \apj, 506, 384

\reference Giavalisco, M., \& Dickinson, M.  2001, \apj, 550, 177

\reference Guarnieri, M. D., Renzini, A., \& Ortolani, S.  1997, \apj, 477, L21

\reference Kauffmann, G., White, S. D. M., \& Guiderdoni, B.  1993, \mnras, 297, L23


\reference McWilliam, A. 1997, \araa, 35, 503

\reference McWilliam, A., \& Rich, R. M.  1994, \apjs, 91, 749

\reference Morris, M., \& Serabyn, G.  1996, Nature, 382, 602

\reference Ortolani, S., Renzini, A., Gilmozzi, R. M., Marconi, G., Barbuy,
B., Bica, E., \& Rich, R. M.  1995, Nature, 377, 701

\reference Peletier, R. F., Balcells, M., Davies, R. L., Andredakis, Y.,
Vazdekis, A., Burkert, A., \& Prada, F.  1999, \mnras, 310, 703


\reference Rich, R. M.  1990, \apj, 362, 604

\reference Rich, R. M. 1999 in The Formation of Galactic Bulges, ed.\ C. M.
Carollo, H. C. Ferguson, \& R. F. G. Wyse (Cambridge: Cambridge University
Press), 54

\reference Rich, R. M., \& McWilliam, A.  2000, in Proc.\ SPIE 4005, Discoveries
and Research Prospects from 8- to 10-Meter-Class Telescopes, ed.\
J. Bergeron (Bellingham: SPIE), 150

\reference Rich, R. M., \& Mould, J.  1991, \aj, 101, 1286

\reference Rich, R. M., Ortolani, S., Bica, E., \& Barbuy, B.  1998, \aj, 116, 1006



\reference Steidel, C. C., Giavalisco, M., Dickinson, M., \& Adelberger,
K. L. 1996, \apj, 462, L17

\reference Stephens, A. W., et al.\ 2001, \aj, 121, 2597

\reference Terndrup, D. M. 1988 \aj, 96, 884

\reference Van den Bergh, S. 1972, \pasp, 84, 306

\reference Van den Bergh, S., \& Herbst, E.  1974, \aj, 79, 603

\reference Vogt, S. S., et al.\ 1994, in Proc.\ SPIE 2198, Instrumentation
in Astronomy VIII, ed.\ D. L. Crawford \& E. R. Craine (Bellingham:
SPIE), 362

\reference Whitford, A. E., \& Rich, R. M.  1983, \apj, 274, 723

\reference Zhao, H., Rich, R. M., \& Spergel, D. N. 1996, \mnras, 282, 175

\reference Zoccali, M., et al.\ 2001, \apj, in press (astro-ph/0101485)

\end{references}
\end{document}